# Sensitivity Enhancement of Pd/Co Bi-layer Film for Hydrogen Gas Sensing using Perpendicular-to-Plane Ferromagnetic Resonance Configuration

Chris Lueng, Peter J. Metaxas, and Mikhail Kostylev

*School of Physics, University of Western Australia, 35 Stirling Highway, Crawley, 6009 WA, Australia.*

**Previously, it has been shown that the strength of the perpendicular magnetic anisotropy (PMA) of thin Palladium-Cobalt bi-layer films can be modified when hydrogen gas is absorbed by the Palladium (Pd) layer. In our recent work we showed that the ferromagnetic resonance (FMR) response of this material is sensitive to changes in PMA upon exposure of Pd to hydrogen gas. As such, a simple, compact and contactless hydrogen gas sensor could exploit FMR-based detection, of the reversible hydrogen-gas-induced changes in PMA. The magnitude of the FMR peak shift is critical in determining the sensor's sensitivity: the higher the FMR peak shifts at a given hydrogen gas concentration, the higher the sensitivity. In the present work we demonstrate that the detection sensitivity is enhanced when the static magnetic field is applied perpendicular to the film plane. A factor of eight times improvement is observed with respect to the in-plane FMR configuration studied previously. An analysis based on the Kittel equation for FMR frequencies of a ferromagnetic film is carried out in order to understand the mechanism of sensitivity enhancement. The result is important for optimizing Pd/Co bi-layered thin films for use in novel platforms for hydrogen gas sensing.**

## I. Introduction

In a number of recent studies [1-7] it has been shown that the strength of the perpendicular interface magnetic anisotropy (PMA) which exists at the interface between a thin layer of ferromagnetic metal (FM) and a layer of palladium (Pd) can be modified if hydrogen gas ($H_2$) is absorbed by the Pd. Pd features reversible absorption of $H_2$, leading to the formation of palladium hydride [8] which is exploited in various hydrogen gas sensor concepts [9].

In our previous work [7] we demonstrated that ferromagnetic resonance (FMR) can be used to detect absorption of $H_2$ by Pd using a Pd/Co bilayer film. Such a detection method may lead to development of a simple contactless and compact hydrogen gas sensor. In the present work we study the sensitivity of the FMR response of these materials to $H_2$ absorption. We find that the FMR peak shift due to $H_2$ absorption is much larger when the static magnetic field is applied perpendicular to the film plane rather than being applied along the film plane. A simple analysis of the Kittel equation is suggested to understand the result.

## II. Experiment

### A. Bi-layer thin film composition

Since the amplitude of the FMR response scales as the volume of the resonating material, it is important to keep the thickness of the cobalt (Co) layer at least a couple of nanometers thick for sensing applications. Therefore, as previously [7,10], we focus on materials with thicknesses ≥5nm. For these materials the strength of PMA is not enough to overcome the film shape anisotropy and the magnetic moment of the Co layer lies naturally in the film plane. For this study, a bi-layer Co(5 nm)/Pd(10 nm) film was sputtered using dc magnetron sputtering in an argon atmosphere onto a Silicon substrate. The base pressure during sputtering was less than 6 mtorr. The sample also underwent in-situ annealing within the chamber immediately after deposition at 200°C for 30 minutes under high vacuum (lower than $10^{-6}$ torr).

### B. Ferromagnetic Resonance Measurements

A gas-tight chamber was used to carry out the FMR measurements [7], the bottom of the chamber being composed of a section of a microwave microstrip line. The microwave ports of the chamber are connected to a microwave generator (used as a source of microwave signal) and to a microwave diode (used to detect the microwave signal passing through the stripline). The rectified signal from the output of the diode is fed into a lock-in amplifier. The chamber is placed between the poles of an electromagnet such that the static field of the magnet is oriented either along the microstrip ("in-plane (IP) configuration") or perpendicular to the sample plane ("perpendicular-to-plane (PP) configuration"). To enable lock-in amplification of the FMR signal, a modulating field of 220 Hz is applied parallel to the static field using small coils with the lock-in amplifier locked to this modulating signal.

The sample under test is placed on top of the microstrip line, with the film facing the line (i.e. film down, substrate up). The onset of FMR absorption results in a drop in the amplitude of the microwave signal at the diode. Since we employ the field modulation method, the signal from the output of the lock-in does not have the shape of a Lorentzian typical for resonances, but the shape of the first derivative of the Lorentzian. In our measurements we keep the microwave frequency constant and sweep the applied field linearly. Accordingly, the registered signal represents the first derivative of the resonance line with respect to the applied field.

## III. Results

### A. Experimental results

Typical examples of the raw FMR traces taken at 12.6 GHz are shown in Fig. 1. The measurements have been taken under two chamber atmospheres, either pure nitrogen gas ($N_2$) or pure hydrogen gas ($H_2$). A reference measurement at ambient atmosphere was also taken before letting $N_2$ into the chamber. No difference between the data taken in air (ambient conditions) and in pure $N_2$ atmosphere was detected. However, a significant resonance field shift was detected when the atmosphere in the chamber was swapped to $H_2$ from $N_2$.

From this figure one sees that the field shifts for IP and PP configurations ($\delta H_{IP}$ and $\delta H_{PP}$ respectively) are in different directions – one observes a downward shift for the field applied IP (Fig. 1(a)) and an upward shift for the PP configuration (Fig. 1(b)). The resonant field shifts and the resonance linewidths extracted from the data in Fig. 1 are given in Table 1.

From Fig. 1 one also sees that $|\delta H_{PP}| \gg |\delta H_{IP}|$. Whereas in the IP configuration the field shift amounts to 50% of the resonance linewidth, in the PP case the resonance is shifted by at least 490% of the resonance linewidth.

### B. Discussion

To understand this behavior we carry out a simple analysis of Kittel Equation for the FMR frequency $f$,

$$f = \gamma\sqrt{(H - H_x + H_z)(H - H_y + H_z)} \quad (1)$$

where $\gamma$ is the gyromagnetic ratio for the material, $H$ is the applied field and $H_x$, $H_y$ and $H_z$ are, respectively, the anisotropy fields along the $x$, $y$ and $z$ axes. The axis $z$ is along the applied field and the static magnetization direction for the material, and $x$ and $y$ are the two directions transverse to the applied field. Equation (1) accounts for both shape and crystalline anisotropy of the sample (including PMA).

Let us first consider the effect of a small change in the transverse anisotropy field (say $H_x$). Solving Eq. (1) for $H$ for a fixed $f$ and differentiating the obtained expression with respect to $H_x$ one obtains

$$S \equiv dH/dH_x = \frac{1}{2} + \frac{H_x - H_y}{2\sqrt{(H_x - H_y)^2 + (2f/\gamma)^2}}. \quad (2)$$

One sees that the sensitivity $S$ strongly depends on the interplay between $H_x$, $H_y$ and $H_z$. For instance, for a very long magnetic rod with circular cross-section magnetized along its length the shape anisotropy fields $H_x = H_y$ and one reaches the maximum sensitivity of $S = 1/2$. For a continuous IP magnetized film with PMA $H_x = -4\pi M_{eff} = -4\pi M + H_{PMA}$, $H_y = 0$ (where $M$ is saturation magnetization and $H_{PMA}$ is the effective PMA field). As a result, Eq.(2) reduces to

$$S_{IP} = \frac{1}{2} - \frac{4\pi M_{eff}}{2\sqrt{(4\pi M_{eff})^2 + (2f/\gamma)^2}}. \quad (3)$$

The second term of Eq. (3) is close to +½ for $H_{PMA} \ll 4\pi M$ and reasonably small frequencies. Hence, the sensitivity $S_{IP}$ is much smaller than ½. The maximum sensitivity in this case is reached for $H_x = 0$ which corresponds to $H_{PMA} = 4\pi M$, i.e. to the threshold of magnetization vector naturally flipping perpendicular to the film plane.

Let us now consider the PP case. In this case, $H_x=H_y=0$ and the absorption of H$_2$ leads to variation in the anisotropy field $H_z$ which is along the applied field, The expression for the sensitivity of the FMR response to slight changes of the longitudinal anisotropy field is much simpler:

$$S_{PP} \equiv dH / dH_z = -1. \qquad (4)$$

The result in Eq. (4) is consistent with the experimental data in Fig. (1). The resonance field shift in the PP configuration is significantly larger than for the IP orientation and it is in the opposite direction.

We checked the validity of this conclusion numerically. The value of $\gamma$=2.88 MHz·Oe$^{-1}$ and $4\pi M_{eff}$=10490 Oe were estimated from the frequency dependence of the PP FMR field for this sample (Fig. 2). IP and PP magnetometry of the saturated film (Figs. 3(a) and (b) respectively) was used to extract a value of $4\pi M \approx 17500$ Oe and the field at which the film is observed to be magnetized perpendicular to the plane via PP magnetometry compares well with the FMR-derived value of $4\pi M_{eff}$=10490 Oe (compare the lateral position of vertical dashed line in Fig. 3(b) to the start of the magnetization plateau). This implies $H_{PMA} = 4\pi M - 4\pi M_{eff} = 7010$ Oe. Most importantly though, on substitution of the $\gamma$ and $4\pi M_{eff}$ values into Eq.(3) we obtained $S_{IP} = 0.116$ giving $|S_{IP}/S_{PP}| = 0.116$. This value is in excellent agreement with the ratio of the experimentally measured peak shifts (Table 1) $|\delta H_{IP} / \delta H_{PP}| = 60/510 = 0.118$.

## IV. CONCLUSION

In conclusion, in this work we demonstrated that the response of the ferromagnetic resonance signal of a Pd/Co bilayer film to hydrogen absorption by its Pd layer is significantly larger for the magnetic static field applied perpendicular to the film plane. This is due to the fact that the sensitivity of the ferromagnetic resonance method to small changes of an effective anisotropy field is larger (or significantly larger, depending on the strength of the anisotropy) for the static magnetic field applied along the anisotropy axis, a result confirmed both via experiment and FMR theory.

## ACKNOWLEDGMENT

This research was supported by the Australian Research Council's Discovery Projects scheme (DP110103980), UWA "Near Miss", "Teaching Relief" and Vice-Chancelor's Research Award schemes and a Small-equipment grant from UWA's Faculty of Science. PJM acknowledges support from the Australian Research Council's Discovery Early Career Researcher Award scheme (DE120100155). The authors acknowledge the instrumentation, and the scientific and technical assistance of the Magnetic Characterisation Facility at The University of Western Australia.

TABLE I
THE FREQUENCY SHIFTS UPON HYDROGENATION UNDER IP AND PP CONFIGURATIONS

| Measurement Configuration | FMR peak AT $N_2$ (Oe) | FMR linewidth AT $N_2$ (Oe) | FMR peak shift upon $H_2$ (Oe) |
|---|---|---|---|
| IP | 1500 | 119 | 60 |
| PP | 15620 | 104 | 510 |

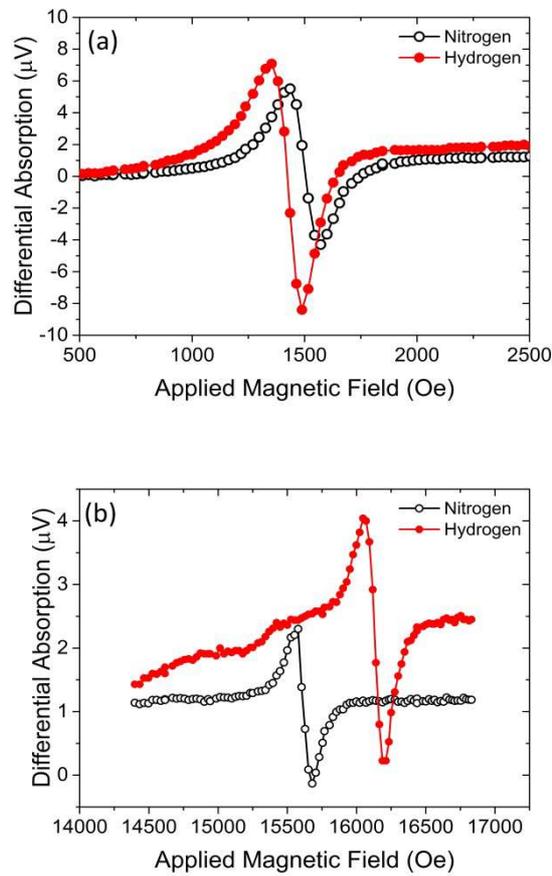

Fig. 1. Typical raw FMR traces taken in pure $N_2$ (black open dots) and pure $H_2$ (red filled dots) gas atmospheres. (a) The static magnetic field is applied in the film plane. (b) The field is applied perpendicular to the film plane. Frequency is 12.6 GHz.

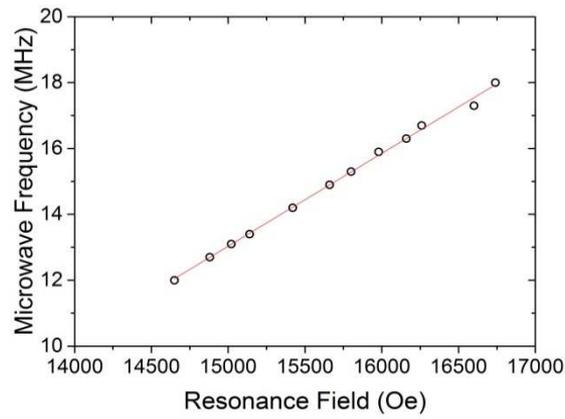

Fig. 2. Frequency dependence of PP FMR field.

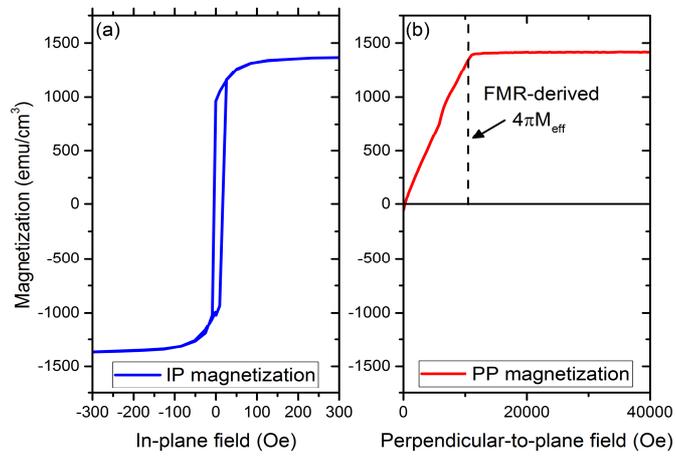

Fig. 3. (a) In-plane magnetization versus in-plane magnetic field and (b) perpendicular-to-plane magnetization versus perpendicular-to-plane magnetic field obtained with a Quantum Design MPMS3.